RESEARCH ARTICLE

# Unveiling the Role of Artificial Intelligence and Stock Market Growth in Achieving Carbon Neutrality in the United States: An ARDL Model Analysis

**Azizul Hakim Rafi[1]\*, Abdullah Al Abrar Chowdhury[1], Adita Sultana[1], Abdulla All Noman[2]**

[1]American National University, 1813 East Main Street, Salem, VA 24153, United States
[2]Montclair State University, Montclair, NJ, 07043, United States

Corresponding author: Azizul Hakim Rafi, Email: rafiazizul96@gmail.com

**Abstract**
Given the fact that climate change has become one of the most pressing problems in many countries in recent years, specialized researches on how to mitigate climate change has been adopted by many countries. Within this discussion, the influence of advanced technologies in achieving carbon neutrality has been discussed. While several studies investigated how AI and Digital innovations could be used to reduce the environmental footprint, the actual influence of AI in reducing $CO_2$ emissions (a proxy measuring carbon footprint) has yet to be investigated. This paper studies the role of advanced technologies in general, and Artificial Intelligence (AI) and ICT use in particular, in advancing carbon neutrality in the United States, between 2021. Secondly, this paper examines how Stock Market Growth, ICT use, Gross Domestic Product (GDP) and Population affect $CO_2$ emissions using the STIRPAT model. After examining stationarity among the variables using variety of unit root tests, this study concluded that there are no unit root problem across all the variables, with a mixed order of integration. The ARDL bounds test for cointegration revealed that variables in this study have a long-run relationship. Moreover, the estimates revealed from ARDL model in the short- and long-run indicated that economic growth, stock market capitalization and population significantly contributed to the carbon emissions in both the short-run and long-run. Conversely, AI and ICT use significantly reduced carbon emissions over both periods. Furthermore, findings were confirmed to be robust using FMOLS, DOLS, and CCR estimations. Furthermore, diagnostic tests indicated the absence of serial correlation, heteroscedasticity and specification errors and, thus, the model was robust.

**Introduction**

Greenhouse gas (GHG) emissions resulting from fossil fuel usage, industrialization, production, forest loss, advances in technology, and growing populations are driving global warming, which is now one of the most serious ecological problems (Nunes,2023; Raihan et al.,2024g). The most advanced nations have prioritized the formulation of strategies to restrict and regulate carbon emissions within their energy and environmental policies (Finon,2019; Raihan et al.,2024e). The USA is acknowledged as a primary contributor to global carbon dioxide ($CO_2$) emissions, significantly impacting the overall concentration of GHG's in the atmosphere (Dogan et al.,2024). Because of its reliance on fossil fuels for energy production, transportation, and industrial processes, the United States is one of the biggest contributors to global carbon emissions. The energy sector is the main source of carbon emissions in the United States, which are mostly caused by power plants, automobiles, and industrial operations (Pata et al.,2023). Notwithstanding this, recent years have seen a trend towards a decrease in emissions, which has been ascribed to a move towards renewable energy sources, energy-saving techniques, and more stringent environmental laws (Kartal, 2023). As of the end of 2021, the USA is the largest economy and the second highest carbon-emitting nation (British Petroleum 2022; World Bank 2022). In 2021, the USA consumed 92.97 exajoules of primary energy and released 4701.1 million tons of $CO_2$ (Kartal, 2023). The selection of the United States as the subject of our research is warranted by its global economic significance and pivotal influence in defining the international landscape. The United States, as the world's largest economy with a GDP of $25.46 trillion in 2022, wields considerable influence (World Bank, 2023). Furthermore, the nation holds the second position in $CO_2$ emissions owing to its rapid economic development and significant energy requirements resulting from its 338 million inhabitants. In 2020, the United States accounted for over 13% of global $CO_2$ emissions, presenting considerable difficulties to its populace and the worldwide biosphere (Adebayo & Ozkan, 2024). In light of these issues, there is a rising consensus on the vital relevance of stock market capitalization, ICT utilization, and AI innovation as strategic remedies. Studies constantly show that progress in AI and ICT can significantly reduce $CO_2$ emissions while promoting the transition to clean, renewable energy. It is expected that further advancements in AI, ICT, GDP, stock market development, and sustainable population growth will reduce environmental costs; therefore, it is imperative to explore the relationship between these factors and $CO_2$ emissions in the United States. The motivation for this research lies in addressing the urgent need to achieve carbon neutrality in the United States amidst growing environmental concerns. Artificial Intelligence (AI) and stock market growth represent transformative forces that can influence sustainable development. AI offers innovative solutions for optimizing energy use, improving industrial efficiency, and promoting green technologies, while stock market growth reflects economic dynamism that can mobilize investments toward low-carbon initiatives.

Climate change has precipitated several catastrophic events, affecting both emerging economies and advanced countries (Shaari et al.,2022; Islam et al.,2023a). In 2023, U.S. energy-related $CO_2$ emissions decreased by 7% compared to 2022, primarily due to a decrease in coal-fired electricity production and a shift towards natural gas and solar energy, primarily from the electric power sector (EIA, 2024). The United States leads the G-7 nations in $CO_2$ emissions, followed by Japan, Germany, Canada, the United Kingdom, Italy, and France (Ayhan et al.,2023). ICT's explosive growth offers nations with new chances to improve their position in the global market, close the gap in economic and social progress, and counteract environmental damage (Asongo et al.,2018, Niebel 2018). The US has seen a significant increase in eco-innovation patents, from 646.32 in 1990 to 4,398 in recent decades, raising concerns about their potential environmental impact (Hossain et al.,2023). A cohort of scholars advocates for the use of ICT communication technology, positing that it contributes to societal education and ultimately plays a part in mitigating carbon dioxide emissions (Ozcan & Apergis 2018; Khan et al.,2020; Shaaban-Nejad & Shirazi, 2022). Moreover, environmental damage, coupled with global disaster, makes up a complex ecological issue (Raihan et al., 2022c) that requires innovative and sophisticated Artificial Intelligence (AI) solutions (Nishant et al., 2020). So, it is essential to emphasize the restricted application of AI in advancing sustainability across sectors such as energy, transportation, water, and biodiversity (Nishant et al., 2020). The public sector's investment in AI has increased significantly in recent decades, exemplified by the $3.2 billion allocated by the U.S. government in 2022 (JEC, 2023).

The United States is at essential intersections, when the need to tackle ecological issues aligns with the pursuit of significant economic progress (Adebayo et al., 2024). In addition, the nation possesses the largest GDP worldwide and allocates substantial funds to its energy infrastructure (Danish & Ulucak, 2021). The World Bank (2020) indicated that in 2018, the United States accounted for around 21.6% of the world GDP (constant 2010 USD).The swift increase of the economy and population, along with the increasing consumption of oil and gas, are the main factors driving the overall trend in developing nations to foster economic expansion (Rahman & Majumder, 2022; Voumik et al., 2023c). The density of individuals in metropolitan regions generates a more extensive labor pool, hence promoting economies of scale and fostering specialization (Raza et al., 2023). This leads to enhanced economic output and productivity (Ridwan,2023). Moreover, renewable energy is often viewed as an exceptionally effective way of advancing environmental sustainability (Onwe et al.,2024; Islam et al.,2024; Ridzuan et al.,2023; Raihan et al.,2024f). Investor perceptions of legal challenges pertaining to potential regulatory dangers and ecological obligations can impact stock market behavior (Topcu et al., 2020). This adverse effect would promote financial decisions that yield superior stock returns and reduced carbon emissions (Mushafiq & Prusak, 2023). Stock market advancements offer investors greater access to funding alternatives, including equity financing, potentially leading to heightened investment in sustainable energy initiatives (Paramati et al. 2016; Sadorsky, 2012).

The USA, despite its substantial contribution to global temperature rise, exhibits a notable study deficiency regarding the effects of multiple variables, rendering it the second-largest global emitter of $CO_2$ (Hassan et al., 2024). This study offers multiple insights into the existing body of work. It represents the inaugural inspection of the interplay between stock market capitalization and AI innovation concerning ecological effects in the United States. Unlike previous studies, this study's methodology allows for differentiation based on the carbon footprints of the USA, not its development levels. Therefore, we will tailor policy suggestions based on the pollution levels of the USA, not its level of advancement. Additionally, the consequence of factors like GDP growth, growing population, and ICT is also categorized based on the total emissions of the country in the evaluation. Incorporating these variables into the empirical model minimizes the risk of omitting key variables. This study is the inaugural complete investigation, within our expertise, of the influence of newly found variables on $CO_2$ emissions, addressing the following principal research questions: What is the impact of AI innovation and stock market capitalization on the environment in the USA? Also, how do ICT utilization, GDP, and growing populations affect carbon intensity in the USA? The study holds significance due to its emphasis on AI innovation and stock market development, areas that previous research has not sufficiently explored. The analysis employed ARDL techniques, utilizing data from 1990 to 2021, and the reliability of the results was further substantiated by FMOLS, DOLS, and CCR techniques. By recognizing these factors, policymakers and strategists can more effectively promote sustainable ethical actions. It provides significant insights for policymakers in the USA and around the world, enabling sustainable revenue growth and enhancing the condition of the planet, particularly through carbon neutrality.

The second part of the inquiry offers an in-depth review of current studies on the chosen determinants. The "Methodology" section fully delineates the data collection procedure, conceptual structure, experimental design of models, and the estimate techniques utilized. The fourth section, headed "Results and Discussion," gives an extensive review of the findings, clarifying the model's consequences. The final section combines the principal findings of the research and offers useful suggestions derived from them.

## Literature Review

Numerous studies have assessed the state of the natural world using various indicators, including $CO_2$ emissions and ecological footprints. We undertook a comprehensive evaluation of the existing academic literature to detect differences. Consequently, we will look at prior research about the influence of $CO_2$ emissions on economic progress, population growth, artificial intelligence (AI), information and communication technology (ICT), and stock market capitalization, which will underpin the requirements of our investigation.

## GDP and CO2 Emission

The correlation between economic progress and green growth has been the focus of numerous researches. For example, Ridwan et al. (2024a) examine the ecological impacts of GDP in six South Asian nations from 1972 to 2021. Utilizing the Driscoll Kraay Standard Error (DKSE) methodology and the CS-ARDL technique, they determined that GDP considerably reduces CO2 emissions in both the short and long term. Similarly, Using the EKC and Pollution Haven Hypothesis (PHH) as a framework, Raihan et al. (2023a) analyse the ecological effects of China's nuclear energy use between 1993 and 2022. The empirical evidence indicated that heightened economic growth could reduce emission levels in the future. Significant growth in economy and abundant resources coincide with heightened ecological degradation (Hunjra et al., 2024). On the other hand, Pattak et al. (2023) elucidate the implications of nuclear, green energy sources, with population and GDP on CO2 emissions in Italy, using the STIRPAT framework from 1972 to 2021. The ARDL paradigm indicates that an increase of 1% in Italian GDP over the long term can result in an 8.08% spike in CO2 emissions. Voumik et al. (2023b) estimate the influence of GDP, population, renewable energy consumption, fossil fuels, and foreign direct investment on Kenya's carbon emissions from 1972 to 2021. Utilizing the ARDL approach, they observed that a boost in Kenya's GDP can elevate the nation's CO2 emissions. Moreover, in their analysis of China's ecological harm, Ahmad et al. (2024a) examines the effects of technology, the economy, and renewable energy. The DOLS estimate indicates that a 1% rise in GDP leads to a 0.51% elevate in CO2 emissions. Multiple studies by Voumik et al.(2023a) in Indonesia, Ridwan et al.(2023) in France, Raihan et al.(2023c) in Malaysia, Raihan et al.(2023b) in Mexico, Rahman et al.(2022) within Bangladesh, Raihan et al.(2022b) in USA and Raihan et al.(2024c) in G-7 region also corroborated with the positive connection between GDP and CO2 emission.

## AI Innovation and CO2 Emission

Artificial intelligence complemented by human skills, carefully assessing AI performance, and accurately defining business goals to ensure the effective alignment of AI technologies (Rahman et al., 2024). From 1990 to 2020, Shiam et al. (2024) look into how innovations in Artificial Intelligence (AI) have affected the Nordic region's ecological footprint. The study takes into account that there is a negative correlation between AI innovation and the ecological footprint, using the STIRPAT model. As per Rasheed et al. (2024), AI plays a proactive role in mitigating carbon emissions while sustaining the ecological balance of seven developing Asian countries. Akther et al. (2024) evaluate the influence of private investment in artificial intelligence (AI) on environmental sustainability in the United States from 1990 to 2019. The findings indicate that private investment in AI significantly correlates with the load capacity factor, hence improving ecological responsibility, as evidenced by the Autoregressive Distributed Lag (ARDL) bound test. The impact of AI innovation on environmental sustainability in the Nordic region is examined by Hossain et al. (2024) between 1990 and 2020. The study used the Panel Autoregressive Distributed Lag (ARDL) model to examine both short-run and long-run interactions, revealing that AI innovation strongly and positively impacts the environment in both time frames. In a similar spirit, Ridwan et al. (2024c) test the Load Capacity Curve (LCC) hypothesis to explore the function of Artificial Intelligence (AI) in fostering sustainability within the G-7 countries. They demonstrate that investing in AI has a major beneficial correlation with the LCF using the Moments Quantile Regression (MMQR) method, hence boosting ecological sustainability. Similar conclusion was also demonstrated by Ridwan et al.(2024b) in USA and Dong et al.(2023) in China. However, Al-Sharafi et al.

(2023) discovered that although AI solutions can save costs, conserve assets, and enhance disposal of waste, their effect on the planet is negligible, especially in developing countries.

## SMC and CO2 Emission

Stock market capitalization (SMC) offers innovative, green technology to nations at all stages of growth, improving energy usage and fostering ethical production to lower $CO_2$ emissions (Piñeiro Chousa et al.,2017; Tanchangya et al.,2024; Ahmad et al.,2024b). In Asian nations, Liang et al. (2023) investigate the impact of energy transition and stock market capitalization on ecological health between 1994 and 2020. The outcomes suggest that SMC can enhance the surrounding conditions. In a similar vein, Musah (2023) explored the relationship between EU environmental quality and the growth of stock markets between 1995 and 2014. According to their findings, the development of the stock market reduced ecological impact and thereby boosted sustainability. Furthermore, Paramati et al.(2017) performed a study in G-20 countries and found that SMC reduces carbon footprint only in emerged countries. Focusing on rapid revenue in the stock market may push companies to prioritize profits over ecological areas, possibly leading to increased harm to the planet (Taghizadeh-Hesary et al. 2022). Zhao et al. (2023) did studies in the BRICS-T countries to explore the link between SMC and emissions of $CO_2$. Between 1990 and 2018, they demonstrated how the development of the stock market leads to a decline in ecological quality using second-generation approaches. Similarly, Zeqiraj et al. (2020) investigated the changing connection between the growth of stock markets and carbon emissions in low-carbon nations from 1980-2016. They established that SMC raises the intensity of carbon emissions over the short and long terms using the CS-ARDL approach. The destructive correlations between the SMC and $CO_2$ emission was also observed by several studies like Shiam et al.(2024) in Nordic area, Zafar et al. (2019) in G-7 zone, Su (2023) in China, Nguyen et al.(2021) in G-6 countries. On the other hand, Azeem et al. (2023) analyzed the influence of stock market capitalization (SMC) on the release of carbon in 40 major carbon-emitting countries from 1996 to 2018. Utilizing the Driscoll-Kraay technique, they discovered an inverted U link between SMC and environmental damage.

## ICT and CO2 Emission

Information and Communication Technology (ICT) significantly influences the environment and has profound implications for prosperity and social growth (Islam & Rahaman,2023).The manufacturing and processing of ICT gadgets is the reason for the degradation of the ecosystem (Danish et al., 2019). A lot of researches have been done regarding the effects of ICT on the surroundings. To determine the precise impact of ICT on harmful emissions, we examine relevant study articles. To determine the effect of ICT on $CO_2$ emissions, You et al. (2024) analyzed panel data from 64 "Belt and Road Initiative economies between 2000 and 2021. Utilizing the Mean Group (MG) estimator, the Augmented Mean Group (AMG) estimator, and the Dumitrescu-Hurlin panel causality, they discovered a reverse connection between $CO_2$ emissions and ICT use. Several examinations also illustrate similar outcomes such as Lu (2018) in 12 asian economies, Batool et al.(2019) in South Korea, Godil et al.(2020) in Pakistan, Appiah-Otoo et al.(2022) in 110 countries, Islam et al.(2023b) in GCC countries, and Tsimisaraka et al.(2023) in OBOR areas. Nevertheless, Uddin et al. (2024) looks into how ICT has affected G20 countries' $CO_2$ emissions between 1980 and 2019. This study confirms the considerable and positive influence of ICT on $CO_2$ emissions using the panel ARDL technique and the Generalized Method of Moments (GMM) calculation. Raihan (2024) examines the impact of ICT on $CO_2$ emissions in Malaysia from 1990 to 2020 by employing the Dynamic Ordinary Least Squares (DOLS) approach. The result demonstrates that the rise in $CO_2$ emissions is affected by ICT utilization. Moreover, Yahyaoui (2024) shows that ICT has a long-term positive effect on $CO_2$ emissions in both Morocco and Tunisia. Additionally, Arshad et al. (2020) assessed the influence of ICT on CO2 emissions across 14 South and Southeast Asian nations from 1990 to 2014. The researchers utilized the PMG, DOLS, and FMLOS techniques and determined that ICT adversely affected environmental quality in the region.

## Population and CO2 Emission

Over the past few decades, population expansion has been a major factor in the rise in worldwide CO2 emissions (Rehman et al.,2022). Due to the increased demand for housing, healthcare, education, and transportation, increasing populations are considered to have a detrimental effect on the environment (Isik et al., 2019; Wu et al., 2021). Hassan et al. (2024) consider the link between nuclear energy, population, and CO2 emissions for the United States. They discovered that population-induced pollution appears in both the short and long-term by using the ARDL simulations model. In five of Asia's most populous regions, Rehman and Rehman (2022) assess the underlying effects of population growth on CO2 emissions between 2001 and 2014. They discovered that population expansion is the most intense component of the CO2 emissions using a gray relational analysis (GRA). Similarly, Khan et al.(2021) checked the association between growth in population on ecosystem in USA from 1971 to 2016. They discovered an encouraging association between growing populations and CO2 emissions and the ecological footprint using the GMM, robust least-squares, and generalized linear model (GLM). However, using rigorous econometric techniques, Pickson et al. (2024) explore population-related factors that affect CO2 emissions from 1993Q1 to 2018Q4, covering a range of income levels in different nations. They revealed that in high and lower-middle-income countries, the density of people reduces CO2 emissions, but in lower-income ones, it increases emissions. In a similar vein, Erdogan (2024) undertook an investigation in Germany from 1995 to 2020. They discovered that population density reduces environmental pollution in Germany by applying the ARDL technique. Moreover, Wu et al. (2021) observed that China's growing population could potentially yield both short and long-term advantages in mitigating biodiversity loss. By comparison, Begum et al. (2015) demonstrated that the population rise does not significantly impact environmental damage in Malaysia, based on the ARDL bounds testing approach.

## Literature Gap

Even if the USA supports sustainable environmental quality, the methodologies for collecting information on ICT use, AI innovation, and the actual impact of stock market capitalization on CO2 emissions are still not well defined. From the vantage point of the USA, domains such as AI innovation, ICT utilization, and stock market evolution remain comparatively underexplored research subjects. Furthermore, our research employs the ARDL limits testing methodology, a technique that has been infrequently utilized in prior investigations. This method facilitates a more efficient examination of data from panel models, which enhances conceptual comprehension in the discipline. By analyzing these characteristics, the chosen area can assess whether technical innovations, economic cooperation, and sustainable development can aid in addressing the planet's sustainability issues. This study addresses a deficiency in the literature by examining the evolving effects of GDP, population growth, ICT utilization, stock market capitalization, and AI on CO2 emissions, employing sophisticated economic methods, with particular emphasis on the United Nations' objectives.

## Methodology

### Data and Variables

The current study examined data pertaining to the impact of specific factors on the environmental quality of the United States from 1990 to 2021. We obtained the Gross Domestic Product (GDP) and demographic data from the World Development Indicators (WDI). In addition, $CO_2$ emissions, utilized as an indicator of ecological sustainability, were likewise obtained from WDI as the endogenous variable. Information regarding AI innovation and ICT utilization was sourced from Our World in Data, whereas stock market capitalization (SMC) statistics were acquired from the Global Financial Development (GFD) database. In this analysis, stock market capitalization and AI innovation were regarded as essential policy variables. Consequently, enhancing the accessibility and trustworthiness of the study's approach ensures that the full documentation provides an explicit and integrated analysis.

**Table 1.** Source and Description of Variables

| Variables | Description | Logarithmic Form | Unit of Measurement | Source |
|---|---|---|---|---|
| CO2 | CO2 Emission | LCO2 | CO2 Emission (kt) | WDI |
| GDP | Gross Domestic Product | LGDP | GDP per capita (current US$) | WDI |
| AI | AI Innovation | LAI | Patent Application in AI field | Our World in Data |
| SMC | Stock Market Capitalization | LSMC | Stock Market Capitalization (% of GDP) | Global Financial Development |
| ICT | Technological Innovation | LICT | ICT good imports (% of total goods imports) | Our World in Data |
| POP | Population | LPOP | Population, total | WDI |

### Theoretical Framework

The IPAT model is a highly focused framework utilized for analyzing the impact of economic activity on energy usage and ecological results (Borsha et al., 2024). This model has been extensively employed in previous studies to examine factors affecting ecological degradation across several contexts (Shaheen et al., 2022; Yu et al.,2023; Wu et al., 2024; Khan, 2024). The model asserts that the environmental impact, represented by the letter "I," is the product of three variables: population (P), affluence (A), and technical advancement (T) (Ehrlich & Holden, 1971).

$$I = \int PAT \qquad (1)$$

This research uses $CO_2$ emissions as a stand-in for environmental decline. In accordance with the STIRPAT model proposed by Dietz and Rosa (1997), we applied population growth as a metric for population (P), economic growth and stock market capitalization as indicators of affluence (A), and the adoption of AI and ICT as a gauge of technology (T). Equation (2) displays the revised form subsequent to the incorporation of the intercept component (C) and the standard error term (ε).

$$I_i = C . P_i^{\beta} . A_i^{\gamma} . T_i^{\delta} . \varepsilon_i \qquad (2)$$

The factual framework established in this paper results from an in-depth look of pertinent research, which has guided the ensuing interpretations.

$$Environmental\ Impact = f(Population, Affluence, Technology) \qquad (3)$$

Alongside independent variables, we incorporated CO2 emissions as a proxy indicator. In this context, GDP refers to gross domestic product, AI denotes artificial intelligence, SMC represents stock market capitalization, ICT means information and communication technology and POP pertains to population. In equation (4), we adjusted $\alpha_1$ to $\alpha_5$ for the coefficients of the independent variables, whereas $\alpha_0$ represents the intercept term. The logarithmic forms of the variables are utilized in equation (5) to guarantee normal distribution. To derive Equation (4), execute the subsequent procedure:

$$CO_{2it} = \alpha_0 + \alpha_1 GDP_{it} + \alpha_2 AI_{it} + \alpha_3 SMC_{it} + \alpha_4 ICT_{it} + \alpha_5 POP_{it} \qquad (4)$$

The logarithmic forms of the variables are utilized in equation (5) to guarantee normal distribution.

$$LCO_{2it} = \alpha_0 + \alpha_1 LGDP_{it} + \alpha_2 LAI_{it} + \alpha_3 LSMC_{it} + \alpha_4 LICT_{it} + \alpha_5 LPOP_{it} \qquad (5)$$

**Estimation Strategies**

This investigation implemented the ARDL approach to analyze the correlation between CO2 emissions and critical variables including GDP, AI innovation, SMC, ICT utilization, and population (POP) in the USA. Initially, we conducted unit root tests (ADF, P-P, and DF-GLS) to establish the stationarity of the variables. The ARDL limits test was used to investigate the cointegration among the variables, considering the peculiarities of the time series data. Additional estimating methods, such as FMOLS, DOLS, and CCR, were applied to guarantee robustness. After a comprehensive evaluation, the most effective and reliable econometric method was selected for the research.

**Unit Root Test**

The researchers conducted unit root analysis to validate the preference for the ARDL methodology over traditional cointegration methods. Utilizing a unit root test is crucial to avert erroneous regression analysis. This testing process determines the degree of integration (Polcyn et al.,2023; Ridwan et al.2024e). In this study, three stationarity tests were run: the DF-GLS test, recommended by Elliot et al. (1996), the Phillips and Perron (1988) test, and the Augmented Dickey-Fuller (ADF) test, which Dickey and Fuller (1981) proposed. In contrast to the Dickey-Fuller (DF) method, the ADF technique is more resilient and suitable for more complex procedures (Fuller, 2009).

**ARDL Bound test**

This study utilized ARDL bound testing (Pesaran et al., 2001) to determine the presence of cointegration among the variables. It is extensively utilized in econometric analyses to examine long-term cointegration among factors and to assess the impact of exogenous variables on the endogenous variable in both the long and short term (Ahmed et al., 2021; Raihan,2023; Atasoy et al.,2022b; Raihan & Bari, 2024).The ARDL limits test is superior to previous single-equation approaches in some respects when examining cointegration (Rahman & Islam, 2020; Ridwan & Hossain, 2024). The ARDL bounds testing methodology is reliable and efficient, even in limited information ranges, providing a thorough evaluation of the overarching structure in the long run. It can be used regardless of the integration order of the fundamental ARDL structure, which can be either of order 2 (I(2)) or order 0 (I(0)) or 1 (I(1)). Equation (6) mathematically displays the ARDL bounds test as follows:

$$\Delta LCO_{2t} = \beta_0 + \beta_1 LCO_{2t-1} + \beta_2 LGDP_{t-1} + \beta_3 LAI_{t-1} + \beta_4 LSMC_{t-1} + \beta_5 LICT_{t-1} + \beta_6 LPOP_{t-1}$$
$$+ \sum_{i=1}^{q} \alpha_1 \Delta LCO_{2t-i} + \sum_{i=1}^{q} \alpha_2 \Delta LGDP_{t-i} + \sum_{i=1}^{q} \alpha_3 \Delta LAI_{t-i} + \sum_{i=1}^{q} \alpha_4 \Delta LSMC_{t-i}$$
$$+ \sum_{i=1}^{q} \alpha_5 \Delta LICT_{t-i} + \sum_{i=1}^{q} \alpha_6 \Delta LPOP_{t-i} + \varepsilon_t$$

(6)

where q is the optimum lag length.

Pesaran et al. (2001) propose that F-statistics may be contrasted with critical values for both upper and lower bounds. If the F-statistics above the upper critical value, the null hypothesis (H0) is rejected, indicating a sustained connection. If the F-statistic falls below the lower critical value, the null hypothesis (H0) is upheld, however its validity remains ambiguous within the specified thresholds.

**ARDL short and long run simulation**

The investigation use the ARDL framework to examine the interaction of variables, taking into account both short-term and long-term dynamics. The long-run coefficient estimate is predicted by equation (7), which also confirms the cointegration of the parameters. It incorporates the ECT into the ARDL framework to compute short-term dynamic parameters derived from long-term estimates, employing an error correction term (ECT) approximation. The equation (7) represents the ARDL long run equation below.

$$\Delta LCO_{2t} = \beta_0 + \beta_1 LCO_{2t-1} + \beta_2 LGDP_{t-1} + \beta_3 LAI_{t-1} + \beta_4 LSMC_{t-1} + \beta_5 LICT_{t-1} + \beta_6 LPOP_{t-1}$$
$$+ \sum_{i=1}^{m} \alpha_1 \Delta LCO_{2t-i} + \sum_{i=1}^{m} \alpha_2 \Delta LGDP_{t-i} + \sum_{i=1}^{m} \alpha_3 \Delta LAI_{t-i} + \sum_{i=1}^{m} \alpha_4 \Delta LSMC_{t-i}$$
$$+ \sum_{i=1}^{m} \alpha_5 \Delta LICT_{t-i} + \sum_{i=1}^{m} \alpha_6 \Delta LPOP_{t-i} + \Omega ECT_{t-1} + \varepsilon_t$$

(7)

where $\Omega$ represents the coefficient of the ECT.

**Robustness Check**

This study employed the Fully Modified Ordinary Least Squares (FMOLS), Dynamic Ordinary Least Squares (DOLS), and Canonical Cointegrating Regression (CCR) methods to judge the reliability of the ARDL findings. These techniques have diverse benefits, as evidenced in the previous research (Merlin & Chen, 2021). Hansen and Phillips (1988) created the FMOLS analysis to integrate the most precise cointegration measures. To deal with the effects of cointegration on serial correlation and endogeneity in the explanatory variables (Zimon et al., 2023), this method changes the least squares method. Furthermore, it can clarify the causal relationships between the factors under study across a broad range of values (Pedroni, 2001). Stock and Watson (1993) derive an ongoing link in an illustration where the elements cointegrate but have a range of integration using parametric approaches in their DOLS framework. The proposed DOLS technique addresses the issues of simultaneity aversion and small sample bias through the use of leads and lags. The primary advantage of this assessment lies in its capacity to illustrate varying degrees of integration among discrete pieces inside the cointegrated framework (Raihan & Tuspekova, 2022). Park (1992) introduced the CCR approach, applicable for identifying cointegrating vectors in a system characterized by an integrated process of order one, denoted as I(1). Furthermore, it is applicable for both single equation regression and multivariate regression without modifications, demonstrating its continued utility.

**Results and Discussion**

Table 2 presents a comprehensive analysis of the variables. It encompasses statistical markers from normality evaluations, such as skewness, probability, kurtosis, and the Jarque-Bera test. The mean and median values for all variables exhibit similarity, indicating a normal distribution. Everything shows an inverse skewness, according to the results, with the exception of AI innovation. The skewness values, approximating 0, signify that all variables conform to a normal distribution. All the series exhibit platykurtic characteristics, with kurtosis values fewer than 3. The Jarque-Bera probability indicates that all variables have a normal distribution.

**Table 2.** Summary Statistics

| Statistic | LCO2 | LGDP | LAI | LSMC | LICT | LPOP |
|---|---|---|---|---|---|---|
| Mean | 15.4644 | 10.6439 | 7.5055 | 4.7704 | 2.6251 | 19.4995 |
| Median | 15.4519 | 10.7189 | 7.1577 | 4.8772 | 2.6312 | 19.5091 |
| Maximum | 15.5692 | 11.1594 | 9.7244 | 5.2724 | 2.8711 | 19.6207 |
| Minimum | 15.2789 | 10.0812 | 6.3208 | 3.9489 | 2.2675 | 19.3355 |
| Std. Dev. | 0.08024 | 0.31878 | 1.0359 | 0.32158 | 0.13253 | 0.08679 |
| Skewness | -0.4666 | -0.2557 | 1.1557 | -0.7628 | -0.2705 | -0.3112 |
| Kurtosis | 2.6354 | 1.8889 | 2.9923 | 2.9142 | 3.7837 | 1.8948 |
| Jarque-Bera | 1.3384 | 1.9948 | 7.1232 | 3.1131 | 1.2092 | 2.145 |
| Probability | 0.5121 | 0.3688 | 0.0284 | 0.2109 | 0.5463 | 0.3422 |

Table 3 illustrates the results of the unit root analysis using the ADF, DF-GLS, and P-P tests. The findings reveal that LICT and LPOP had stationary behavior at both the level and initial difference, as evidenced by the ADF, P-P, and DF-GLS tests. However, these tests identified the other variables (LCO2, LGDP, LAI, and LSMC) as non-stationary at I(0) and attained stationarity at I(1). The results of these tests necessitate the implementation of the study using the ARDL approach.

**Table 3.** Results of Unit root test

| Variables | ADF | | P-P | | DF-GLS | | Decision |
|---|---|---|---|---|---|---|---|
| | I(0) | I(1) | I(0) | I(1) | I(0) | I(1) | |
| $LCO_2$ | -0.155 | -4.954*** | -0.231 | -4.356*** | -0.221 | -3.889*** | I(1) |
| LGDP | -0.878 | -4.672*** | -0.762 | -4.032*** | -0.760 | -4.140*** | I(1) |
| LAI | -0.258 | -3.981*** | -0.316 | -4.002*** | -0.289 | -4.081*** | I(1) |
| LSMC | -0.416 | -4.990*** | -0.336 | -4.821*** | -0.435 | -4.779*** | I(1) |
| LICT | -3.061** | -4.585*** | -3.981*** | -4.550*** | -3.450** | -4.089*** | I(0) |
| LPOP | -5.088*** | -6.451*** | -4.827*** | -6.778*** | -5.064*** | -6.566*** | I(0) |

The ARDL bound test results, presented in Table 4, provide significant information about the cointegration of the components under study. The F-statistic value of 5.86, exceeding the upper limits for significance at the 10%, 5%, 2.5%, and 1% levels for both zero and first orders, suggests that H0 is false. This observation reveals a persistent link between the variables.

**Table 4.** Results of ARDL Bound test

| | Test Statistics | Value | K | |
|---|---|---|---|---|
| | F statistics | 5.8605 | 5 | |
| | Significance level | | | |
| Critical Bounds | 10% | 5% | 2.50% | 1% |
| I(0) | 2.08 | 2.39 | 2.70 | 3.06 |
| I(1) | 3 | 3.38 | 3.73 | 4.15 |

Table 5 presents the entire set of the ARDL simulation results. In the short term, CO2 emissions increase by 0.161% for each 1% rise in GDP. Over time, a 1% increase in GDP intensifies the correlation, resulting in a 0.354% rise in carbon emissions. The likely explanation for this phenomenon is that increased economic activity typically leads to increased energy consumption and industrial production, which often rely on fossil fuels, thereby causing environmental degradation. Several studies have corroborated with this findings such as Raihan et al.(2024d) in Indonesia, Sun et al.(2024) in 17 APEC countries, Raihan et al.(2024b) in Vietnam, Cao et al.(2022) in OECD economies, Abid et al. (2022) In G-8 countries, Pata et al.(2023) in USA, Raihan et al.(2024h) within Bangladesh, Mehmood (2024) in South Asian countries and Chen et al. (2022) in BRICS zone. On the other hand, Raihan et al.(2024a) in India and Saqib and Usaman (2023) in USA explained that economic growth is beneficial for the ecosystem level. Moreover, Salari et al. (2021) found an inverted-U shape relationship between $CO_2$ emissions and GDP in USA.

Conversely, a 1% increase in AI innovation correlates with a short-term reduction of 0.053% and a long-term decrease of 0.113% in CO2 emissions. AI innovation may enhance the natural environment by maximizing resource utilization, increasing energy efficiency, and advocating for sustainable behaviors across various industries. This result is aligns with Liu et al.(2022), Wang et al.(2023), Ding et al. (2023), Ahmad et al.(2021), Abir et al.(2024), and Bala et al.(2024). Conversely, Nahar (2024) indicated that AI-driven innovation did not have a beneficial impact on SDGs 10, 12, and 14–15 for the majority of nations among the 22 examined. On the other hand, a 1% increase in stock market value results in an immediate rise of 0.125% in carbon emissions and a long-term increase of 0.177%. The capitalization of the stock market elevates carbon emissions because heightened market activity frequently stimulates industrial growth and production, resulting in greater energy consumption and CO2 emissions. Similar outcome was also demonstrated by Alam et al.(2021); Ridwan et al.(2024d) causes more carbon emission.

Conversely, a 1% increase in ICT adoption results in a reduction of carbon emissions by 0.092% in the short term and 0.578% in the long term. We can assume that enhancing resource management efficiency, facilitating remote work, and promoting cleaner technology can make ICT consumption more environmentally friendly. Our findings is supported by (Danish,2019; Usman et al.,2021; Raihan et al.,2022a; Atasoy et al.,2022a; Qayyum et al.,2024) concluded that ICT might be utilized to lessen the detrimental effects of CO2 emissions and enhance the environmental quality. Conversely, Raheem et al.(2020), Huang et al.(2022) in E-7 economy claimed that the elevation in emissions causes by ICT use and hampers environment sustainability. A 1% increase in population growth leads to a 0.952% rise in CO2 emissions in the short term and a 0.810% increase in the long run. This may occur owing to heightened demand for energy, transportation, and resources, resulting in greater fossil fuel use and waste generation. In a similar vein, Voumik and Ridwan (2023) in Argentina, Voumik et al.(2023b) in Kenya and Appiah et al. (2023) in OECD economies expressed that population growth is harmful for the ecosystem. P-values below the established 5% significance level indicate that the coefficients demonstrate statistical significance in both time frames. These findings demonstrate that in the USA, there exists an inverse association of GDP, SMC, and population growth on carbon emissions. Conversely, studies have noted that the application of ICT and AI innovations reduces CO2 emissions in both the short and long term.

**Table 5.** Results of ARDL short-run and long-run Estimation

| Variable | Coefficient | Std. Error | t-Statistic | Prob. |
|---|---|---|---|---|
| Long-run Estimation | | | | |
| LGDP | 0.354 | 1.503 | 1.897 | 0.020 |
| LAI | -0.113 | 0.049 | -2.303 | 0.027 |
| LSMC | 0.177 | 0.337 | 0.524 | 0.043 |
| LICT | -0.578 | 0.464 | -1.246 | 0.003 |
| LPOP | 0.810 | 0.034 | 0.756 | 0.034 |
| C | 75.201 | 13.863 | 0.907 | 0.015 |
| Short-run Estimation | | | | |
| D(LGDP) | 0.161 | 0.177 | 2.153 | 0.000 |
| D(LAI) | -0.053 | 0.031 | -1.713 | 0.002 |
| D(LSMC) | 0.125 | 0.027 | 0.455 | 0.003 |
| D(LICT) | -0.092 | 0.071 | -1.196 | 0.051 |
| D(LPOP) | 0.952 | 1.424 | 3.476 | 0.013 |
| CointEq(-1)* | -0.226 | 0.038 | -5.891 | 0.000 |

Table 6 shows the results of the robustness investigation. It shows that the FMOLS, DOLS, and CCR simulations consistently showed good performance, producing results similar to those produced by the extended ARDL estimates over time. The GDP coefficients in all three models (FMOLS, DOLS, and CCR) are statistically significant at the 1% level and display positive values. A 1% increase in GDP results in a corresponding rise in LCO2 emissions of 0.977%, 0.872%, and 0.652% across the models, respectively. The FMOLS model suggests that a 1% increase in the LAI coefficient results in a 0.101% decrease in LCO2;

however, this finding is statistically insignificant. In the DOLS and CCR models, a 1% increase in LAI results in a considerable reduction of LCO2 by 0.192% and 0.017%, respectively. Furthermore, a 1% increase in stock market capitalization (LSMC) results in a 0.032% rise in LCO2 emissions in the FMOLS model, but a 1% increase in ICT usage (LICT) leads to a 0.250% decrease in emissions. These results align with the ARDL short-run and long-run estimations. A 1% increase in LPOP leads to a 0.342% increase in LCO2 emissions, as per FMOLS analysis. According to the DOLS model, a 1% rise in LSMC and LPOP results in an increase of LCO2 emissions by 0.081% and 0.256%, respectively. A 1% increase in ICT enhances environmental quality by decreasing LCO2 emissions by 0.186%, a finding that is statistically significant at the 1% level. In the CCR model, a 1% increase in LSMC and LPOP results in a 0.076% and 0.231% increase in LCO2 emissions, respectively, but a 1% increase in ICT leads to a 0.234% decrease in LCO2 emissions. The results align with the ARDL findings presented in Table 5, thereby strengthening the analysis's robustness.

**Table 6.** Results of Robustness Check

| Variables | FMOLS | DOLS | CCR |
|-----------|-------|------|-----|
| LGDP | 0.977*** | 0.872*** | 0.652*** |
| LAI | -0.101 | -0.192*** | -0.017*** |
| LSMC | 0.032** | 0.081** | 0.076** |
| LICT | -0.250** | -0.186*** | -0.234** |
| LPOP | 0.342** | 0.256*** | 0.231*** |
| C | 62.915*** | 59.566*** | 61.789*** |

Furthermore, this study included a number of diagnostic procedures to check how accurate the ARDL results were. Table 7 contains the calculations for the diagnostic assessment of the ARDL approach. The model functioned without any faults. The Jarque-Bera test, yielding a p-value of 0.57129, suggests that the residuals follow a normal distribution. The Lagrange multiplier analysis indicates the absence of serial correlation in the residuals, with a p-value of 0.06712. The Breusch-Pagan-Godfrey test indicates that the residuals do not display heteroscedasticity, as evidenced by a p-value of 0.3411. The diagnostic procedures applied to the ARDL model showed a high degree of agreement.

**Table 7.** Results of Diagnostic Test

| Diagnostic tests | Coefficient | p-value | Decision |
|------------------|-------------|---------|----------|
| Jarque-Bera test | 0.57129 | 0.4031 | Residuals are normally distributed |
| Lagrange Multiplier test | 0.06712 | 0.1023 | No serial correlation exits |
| Breusch-Pagan-Godfrey test | 1.76921 | 0.3412 | No heteroscedasticity exists |

Finally, the findings of causal linkages across several economic indicators are presented in Table 8. An F-statistic of 4.95374 and a p-value of 0.0154 indicate that LLGDP does not Granger-cause LCO2. This suggests that we reject the null hypothesis that there is no link between variables at the 5% significance level. Furthermore, the presence of one-way causation from LAI, LSMC, LICT and LPOP to LCO2 is confirmed by the p-values that are less than the conventional significance threshold. Thus, we rule out the null hypothesis that there is no causal relationship under these circumstances. On the other hand, p-values greater than the traditional significance criterion for each case show that there is no meaningful causal connection from LCO2 to LGDP, LAI, LSMC, LICT and LPOP. These results imply that changes in LCO2 do not influence ICT usage, economic growth, artificial intelligence, population growth and stock market capitalization. So, it is not possible to rule out the null hypothesis that there is no causality in these interactions.

**Table 8.** Results of Pairwise Granger Causality test

| Null Hypothesis | Obs | F-Statistic | Prob. |
|---|---|---|---|
| LGDP $\neq$ LCO2 | | 4.95374 | 0.0154 |
| LCO2 $\neq$ LGDP | 30 | 0.84545 | 0.4413 |
| LAI $\neq$ LCO2 | | 5.58219 | 0.0099 |
| LCO2 $\neq$ LAI | 30 | 1.83762 | 0.1801 |
| LSMC $\neq$ LCO2 | | 8.65988 | 0.0014 |
| LCO2 $\neq$ LSMC | 30 | 1.67744 | 0.2072 |
| LICT $\neq$ LCO2 | | 10.7905 | 0.0004 |
| LCO2 $\neq$ LICT | 30 | 0.0616 | 0.9404 |
| LPOP $\neq$ LCO2 | | 4.60632 | 0.0198 |
| LCO2 $\neq$ LPOP | 30 | 0.63614 | 0.5377 |

**Conclusion and Policy Implications**

This study seeks to analyze the long- and short-term effects of population increase, economic development, AI innovation, stock market capitalization, and ICT utilization on the carbon footprint in the United States, utilizing data from 1990 to 2021. The current research employed the ADF, DF-GLS, and P-P unit root tests to ascertain the integration order of the dataset. The variables exhibited long-term cointegration, as evidenced by the ARDL bounds test. Population growth, stock market development, and economic expansion would exacerbate environmental deterioration in the selected area, whereas advancements in artificial intelligence (AI) and the application of information and communication technology (ICT) would enhance the environment by reducing $CO_2$ emissions. The anticipated results are solid and validated based on the CCR, FMOLS, and DOLS estimators. The Granger causality test suggests that LAI, LSMC, LICT, and LPOP may contribute to the carbon intensity of the USA in relation to LCO2. The diagnostic test confirms the appropriate distribution of the analysis residuals by revealing the absence of autocorrelation and heteroscedasticity. This article presents additional policy ideas for mitigating pollution while promoting sustainable development through the financing of green ICT, equitable advancement, sustainable stock market practices, and increased application of AI innovation. Finally, to avert resource depletion and promote sustainable development, the government ought to provide incentives for individuals to use green AI innovations and cutting-edge information technology. This study's findings yield various policy recommendations to improve carbon neutrality initiatives in the United States. The findings indicate that Artificial Intelligence (AI) and Information and Communication Technology (ICT) substantially decrease carbon emissions in both the short and long term, but economic development, population growth, and stock market expansion lead to increased emissions. Policymakers must prioritise the incorporation of AI and ICT in industries by offering incentives for enterprises to implement AI-driven solutions that enhance energy efficiency, minimise waste, and improve environmental performance. Advancing research and development in artificial intelligence for environmental sustainability is essential. Efforts must concurrently focus on dissociating economic growth from carbon emissions by advancing cleaner technologies, sustainable practices, and carbon-neutral regulations in industries such as manufacturing and energy generation. Urban planning and population management strategies must prioritise minimising the environmental impact of increasing populations via sustainable infrastructure and intelligent urban solutions. Ultimately, financial market regulations ought to promote environmentally sustainable investments, guaranteeing that stock market expansion bolsters green technologies and sustainable businesses. These steps can collectively guarantee that economic expansion and population growth do not impede efforts towards carbon neutrality.

*Declaration*

**Acknowledgment:** N/A

**Funding:** N/A

**Conflict of interest:** N/A

**Ethics approval/declaration:** N/A

**Consent to participate:** N/A

**Consent for publication:** N/A

**Data availability:** Available on request

**Authors contribution:** Azizul Hakim Rafi, Abdullah Al Abrar Chowdhury, and Adita Sultana from contributed equally to the conceptualization, data collection, and analysis of the study. Abdulla All Noman from provided support in data interpretation and contributed to the manuscript's critical revision. All authors reviewed and approved the final manuscript.